# Engineering band structures of two-dimensional materials with remote moiré ferroelectricity

Jing Ding[1,2,3], Hanxiao Xiang[1,2,3], Wenqiang Zhou[2,3], Naitian Liu[1,2,3], Xinjie Fang[1,2,3], Kangyu Wang[1,2,3], Linfeng Wu[1,2,3], Kenji Watanabe[4], Takashi Taniguchi[5], Shuigang Xu[2,3*]

[1] *Department of Physics, Fudan University, Shanghai, 200433, China*
[2] *Key Laboratory for Quantum Materials of Zhejiang Province, Department of Physics, School of Science, Westlake University, 18 Shilongshan Road, Hangzhou 310024, Zhejiang Province, China*
[3] *Institute of Natural Sciences, Westlake Institute for Advanced Study, 18 Shilongshan Road, Hangzhou 310024, Zhejiang Province, China*
[4] *Research Center for Electronic and Optical Materials, National Institute for Materials Science, 1-1 Namiki, Tsukuba 305-0044, Japan*
[5] *Research Center for Materials Nanoarchitectonics, National Institute for Materials Science, 1-1 Namiki, Tsukuba 305-0044, Japan*

[*]Corresponding author: xushuigang@westlake.edu.cn

**Abstract:**
The stacking order and twist angle provide abundant opportunities for engineering band structures of two-dimensional materials, including the formation of moiré bands, flat bands, and topologically nontrivial bands. The inversion symmetry breaking in rhombohedral-stacked transitional metal dichalcogenides (TMDCs) endows them with an interfacial ferroelectricity associated with an out-of-plane electric polarization. By utilizing twist angle as a knob to construct rhombohedral-stacked TMDCs, antiferroelectric domain networks with alternating out-of-plane polarization can be generated. Here, we demonstrate that such spatially periodic ferroelectric polarizations in parallel-stacked twisted $WSe_2$ can imprint their moiré potential onto a remote bilayer graphene. This remote moiré potential gives rise to pronounced satellite resistance peaks besides the charge-neutrality point in graphene, which are tunable by the twist angle of $WSe_2$. Our observations of ferroelectric hysteresis at finite displacement fields suggest the moiré is delivered by a long-range electrostatic potential. The constructed superlattices by moiré ferroelectricity represent a highly flexible approach, as they involve the separation of the moiré construction layer from the electronic transport layer. This remote moiré is identified as a weak potential and can coexist with conventional moiré. Our results offer a comprehensive strategy for engineering band structures and properties of two-dimensional materials by utilizing moiré ferroelectricity.

**Main text:**

Engineering band structures of quantum materials offers an efficient way to alter their physical properties and to explore emergent phenomena. In two-dimensional materials, controlling the twist angle or stacking order during the van der Waals assembly can create versatile interfacial structures while maintaining clean interfaces[1-9]. Designer van der Waals heterostructures can remarkably alter the electronic band structure of their parent compound, resulting in a plethora of exotic physical phenomena[10]. Particularly, moiré superlattices, constructed by stacking two sheets of layered materials with a small twist angle or lattice mismatch, have been already demonstrated as a powerful platform for studying Hofstadter's butterfly[11-13], strong correlations[14], superconductivity[15], orbital magnetism[16], ferroelectricity[17], and topological states[18,19]. Nevertheless, current widely used technologies for constructing moiré superlattices require the constituent materials to function as both moiré construction layers and electronic transport layers, thus limiting the range of applicable materials. Additionally, these moiré superlattices typically exhibit short-range interactions arising from interlayer hybridizations or atomic reconstructions.

The stacking order plays a particularly significant role in inversion symmetry broken systems, as evidenced by recent discoveries of interfacial ferroelectricity in rhombohedral- (parallel-) stacked h-BN[20-22], rhombohedral-stacked transitional metal dichalcogenides (TMDCs)[23-26], and similar materials[27,28]. The symmetry breaking with the absence of an inversion center leads to an out-of-plane electrical polarization, the direction of which depends on the stacking order of the materials[29]. A moiré superlattice formed by parallelly twisted TMDCs exhibits periodic local atomic registrations, resulting in alternating MX/XM stacking configurations with sub-micron sizes (Fig. 1b and 1c), where M and X represent the metal and chalcogen atoms, respectively. This configuration yields an array of polar domains in the moiré length scale with anti-aligned dipoles[26,30,31]. The built-in moiré ferroelectricity in these systems can generate a triangular superlattice potential in adjacent environments (Fig. 1e), characterized by long-range interactions and noninvasiveness[32-36]. Therefore, one can selectively choose a target material to experience such moiré ferroelectric potential and flexibly engineer band structures to explore exotic electronic properties.

Here, we report a method for constructing remote moiré superlattices by employing both twist angle and stacking order. In our approach, the moiré construction layer is separated from the target layer responsible for the electronic transport. As a demonstration, we utilized twisted $WSe_2$ with a rhombohedral-stacked interface as the moiré construction layer, with its moiré period controllable via the twist angle of $WSe_2$. Bilayer graphene was selected as the target layer, facilitating the high carrier mobility in the transport. As one of the representatives of TMDCs, $WSe_2$ has been demonstrated as a high-quality insulator and high-performance substrate for graphene, given that the Dirac point of graphene resides deep within the band gap of $WSe_2$[37,38]. We observe that the moiré ferroelectric potential present in twisted $WSe_2$ can be imprinted onto bilayer graphene, manifesting as pronounced satellite resistance peaks that are tunable by the twist angle of WSe$_2$.

**Device structure**
To this end, we fabricated a rhombohedral-stacked interface by twisting double bilayer WSe$_2$ to a near 60° angle. A perfect 60° twist in double bilayer WSe$_2$ results in rhombohedral stackings with two possible out-of-plane polarization at the interface, as depicted in Fig. 1c. However, even a slight deviation from this angle can lead to the formation of a moiré superlattice at the interface, characterized by a triangle network, as illustrated in Fig. 1b. We chose double bilayer WSe$_2$ as the building block due to its ability to demonstrate that remote moiré potential can be imprinted onto a target layer without direct contact. Notably, a prominent moiré effect was observed even at a separation distance of 1.4 nm, corresponding to the thickness of the bilayer WSe$_2$.

We targeted a twisted angle at $\theta + 60°$ with $|\theta| \approx 0.5° - 2°$. On one hand, it has been reported that in twisted TMDCs, when the twisted angle is smaller than 2°, significant lattice reconstruction occurs. This reconstruction shrinks the area of AA stacking configuration while increasing the area of MX/XM regions, resulting in a triangular domain structure with narrow domain walls[39]. On the other hand, we aim to avoid the marginally twisted angle region, as distinguishing satellite points in transport measurements becomes challenging due to disorder induced broadening[23,24]. Meanwhile, the angle inhomogeneity formed during the sample fabrication can readily disrupt the regular periodicity. The presence of triangle domain networks in parallelly twisted TMDCs is clearly observed in the piezoresponse force microscopy (PFM) images shown in Fig. 1d. Notably, we observe a significant contrast between the adjacent MX and XM local domains in the PFM images, suggesting that this contrast arises from the piezoelectric or electrostatic effects rather than the flexoelectric effect[40]. In this system, the out-of-plane polarizations alternate in the moiré length scale, resulting in the moiré domain antiferroelectric arrangement[26].

**Remote moiré superlattice**
We utilized an h-BN encapsulation structure to ensure high-quality devices, as illustrated in Fig. 1a. The stack consists of h-BN/bilayer graphene/twisted double bilayer WSe$_2$/h-BN/graphite heterostructures. A dual-gate configuration was employed to independently tune carrier densities $n$ and displacement fields $D$ of our devices. The presence of a moiré superlattice in twisted WSe$_2$ significantly modifies the transport behavior of the remote bilayer graphene. Fig. 1f shows the typical carrier density dependence of the longitudinal resistance $R_{xx}$ for samples with various twist angles. Two symmetric satellite resistance peaks with respect to the charge-neutrality point (CNP) can be observed in all samples. The electrons in bilayer graphene experience a moiré electrostatic potential generated in the parallelly twisted double bilayer WSe$_2$ via ferroelectric polarization, resulting in the formation of a moiré miniband by zone folding. The twist angle $\theta$ of WSe$_2$ gives rise to a moiré pattern with a long wavelength $\lambda$ described by $\lambda = a/2\sin(\theta/2)$, where $a$ is the in-plane lattice constant of WSe$_2$. The carrier density at the full filling of a moiré band in graphene is $n_s = 4/A$, where $A = \sqrt{3}\lambda^2/2$ is the area of a moiré unit cell. The pre-factor of 4 in $n_s$ arises from the four-fold degeneracy in the graphene band structure, which includes two-fold spin and two-fold valley degeneracy.

In Device D1, we intentionally misaligned the graphene with the top h-BN to avoid the formation of additional moiré superlattices. In this device, we observe the satellite peaks at $n_s = \pm 4.9 \times 10^{11}$ cm$^{-2}$, corresponding to a twist angle of $|\theta| = 0.61°$ and a moiré wavelength of $\lambda = 30.8$ nm. The obtained $\lambda$ is much larger than 15 nm, excluding the possibility of the moiré pattern originating from graphene/h-BN interface. This is because the lattice mismatch between graphene and h-BN sets the maximum moiré wavelength at approximately 15 nm, which occurs at zero twist angle.

Figure 2c shows a color plot of the longitudinal resistance $R_{xx}$ as functions of $n$ and $D$. Besides the highly resistive peaks at CNP, two vertical peaks appear at fixed $n_s = \pm 4.9 \times 10^{11}$ cm$^{-2}$. Their positions are independent of $D$, although they gradually merged with the CNP peak at high $D$. To examine the strength of the band folding effect, we measured the temperature dependence of the satellite peaks, as plotted in Fig. 2b. We find that the resistance at $n = 0$ increases with decreasing temperature, demonstrating a typical insulating behavior of bilayer graphene at its CNP. However, the satellite peaks weaken with increasing temperature, disappearing around $T = 50$ K, and show a typical metallic behavior in their temperature-dependent resistances. We further performed the magneto-transport measurements. Figure 2d shows the Landau fan diagram at $D = 0$ V nm$^{-1}$, plotted as $R_{xx}$ as a function of $n$ and magnetic field $B$. Under magnetic fields, the satellite peaks develop into much more resistive peaks, intersecting with the Landau fan from CNP. We can observe some faint Landau level features ($\nu = -4, 4, 8$) fanning out from $n_s$. However, other fractal features, such as Hofstadter's butterfly and Brown-Zak oscillations, are absent in this sample[12]. In general, the moiré effect in our structure exhibits weak interaction.

**Ferroelectric domain switching**
The imprinted moiré superlattice may have several potential origins. First, we exclude the origin of lattice deformation similar to that in graphene/h-BN superlattice[41]. We intentionally chose bilayer WSe$_2$ as the building block to keep the graphene layer at least 1.4 nm away from the moiré interface, thus avoiding direct contact. Remote Coulomb interaction could also contribute to the moiré potential. However, previous studies have shown that efficient electrostatic screening is only effective in a distance of about 1 nm for a typical h-BN spacer with a dielectric permittivity of $\varepsilon \approx 3.5$ [42]. The relatively high dielectric permittivity of WSe$_2$ ($\varepsilon \approx 7$)[37] and the parabolic band structure of bilayer graphene in our structures further reduce the effective distance[42]. Therefore, the remaining origin is the periodic electrostatic potential arising from alternating out-of-plane polarization in twisted WSe$_2$. This can be confirmed by examining the ferroelectric hysteresis in our samples.

Figure 3a shows the enlarged feature near the CNP. Distinguishable split peaks near the CNP can be observed at nonzero $D$ and smear at high temperature (Fig. 3b). The split CNP exhibits hysteresis when we sweep the back gate ($V_b$) forward and backward at a fixed top gate ($V_t = 2$ V) as shown in Fig. 3c. By contrast, when we sweep $V_t$ forward and backward at a fixed $V_b = -1.3$ V, there is no obvious hysteresis. These features have been demonstrated as a hallmark of ferroelectric hysteresis arising from electric field-induced domain switching[23]. Figure 3e shows how the strength of hysteresis depends on $D$ by measuring the difference in $R_{xx}$ between forward and backward sweeps of $V_b$ at each fixed $V_t$. The out-of-plane polarizations in MX/XM domains in parallelly twisted WSe2 induced additional charge carrier in graphene, resulting in a slightly shift in the CNP. At $D = 0$, the two domains induced equal-density carriers of opposite types, resulting in no change in total $n$ originating from the polarization of twisted WSe2, and the CNP of graphene appears as a normal shape. The application of finite $D$ breaks the balance of MX and XM domain areas through the motion of domain walls[23,24,26], causing the CNP to split into two peaks with unequal resistance heights. Moreover, slight motion of domain walls can switch the direction of polarization in either the MX or XM domain by sliding ferroelectricity, leading to hysteresis in $R_{xx}$. These ferroelectric switching behaviors resemble those in marginally twisted TMDCs[23]. The key difference between our samples and previously reported ones is that the twist angle in our Device D1 is much larger, rather than marginal. This allows us to achieve a more uniform moiré structure, thus to observe the band folding effect.

**Coexistence of two types of moiré superlattices**

To compare the superlattices constructed by moiré ferroelectricity (F-moiré) with conventional moiré (C-moiré) superlattices, we fabricated a device (Device D2) containing both types of moirés. The stack was realized by utilizing a similar structure to that shown in Fig. 1a, except that we intentionally aligned the bilayer graphene with the top h-BN and twisted bilayer WSe2 with parallel stacking (target angle at $0° + 0.7°$) was employed in Device D2. The crystallographic alignment was confirmed by Raman spectroscopy (see Fig. S2). The small lattice mismatch between graphene and h-BN produces a C-moiré superlattice on the scale of around 10 nm, where graphene functions as both the moiré construction layer and the electronic transport layer.

Figure 4a and 4b show the $R_{xx}$ as a function of $n$ at $D = 0$ V nm$^{-1}$ and the $n - D$ map, respectively. Compared with the devices shown in Fig. 1f, Device D2 exhibits two sets of satellite resistance peaks. The peaks appearing at $n_s = \pm 6.5 \times 10^{11}$ cm$^{-2}$, close to the CNP, are attributed to F-moiré with a wavelength of $\lambda = 26.6$ nm, while the ones at $n_s' = \pm 3.46 \times 10^{12}$ cm$^{-2}$ arise from C-moiré with a wavelength of $\lambda = 11.5$ nm. The calculated twist angles are 0.71° for twisted bilayer WSe2 and 0.81° for the graphene/h-BN heterostructure. Since the two sets of moirés originate from two separate interfaces, this configuration prevents the formation of super moiré[43]. The independent moirés allow us to *in-situ* compare their influence on the band structure of graphene.

Although we observed similar heights of the satellite resistance peaks corresponding to F-moiré and C-moiré at $B = 0$ T, as shown in Fig. 4a and 4b, their magneto-transport behaviors differ significantly. Figure 4c and 4d display $R_{xx}$ and the corresponding $R_{xy}$ for Device D2 as a function of $n$ and $B$ at $D = 0$ V nm$^{-1}$. Prominent Landau fans emerge from the CNP and $n'_s$, arising from C-moiré. Their intersections form Hofstadter's fractal structures and Brown-Zak oscillations. From the period in $1/B$ of the Brown-Zak oscillation, we can alternatively calculate the twist angle between graphene and h-BN according to $\frac{\phi}{\phi_0} = 1/q$, where $\phi = BA$ is the magnetic flux per moiré unit cell, $\phi_0 = h/e$ is the flux quantum, and $q$ is an integer number. The result is 0.87°, which agrees well with the one calculated from $n'_s$. In comparison, F-moiré exhibits a weaker magnetic response, manifesting as positive magnetoresistance accompanied by faint sign reversal near $n_s$. No prominent Brown-Zak oscillation associated with F-moiré is observed in this device.

**Discussion**

Although the remote moiré interactions appear weak, they hold important potential for modifying the band structure or the topology of two-dimensional materials. For example, as recently reported in rhombohedral pentalayer graphene/h-BN superlattices, a series of fractional quantum anomalous Hall states emerge at the interface away from moiré superlattices[18]. To observe these exotic topological states, a moiré potential is necessary but must be weak enough. It has been proposed that remote moiré superlattices with weak potential can stabilize quantum anomalous Hall crystal and fractional quantum anomalous Hall phases[44-46]. Our strategy offers a promising approach to study correlated topological states with controllable moiré wavelengths in such systems. The period of the moiré potential can be easily tuned by adjusting the twist angle of the remote TMDCs, while its strength can be controlled by selecting different thicknesses of TMDCs as the building blocks. This allows the distance between the target layer and the moiré construction layer to be precisely adjusted.

In conclusion, we have designed a novel heterostructure to create moiré superlattice by imprinting a moiré ferroelectric potential onto a target electronic material. The separation of the moiré potential from the electronic transport layer allows us to preserve high quality in the target layer while designing diverse moiré superlattices with various symmetries. This technology can be easily applied to other polar 2D insulators, such as twisted h-BN with rhombohedral-stacked interfaces, and to arbitrary conducting target materials beyond graphene. Our work provides an avenue to extend moiré superlattice and construct exotic band structures, such as topological flat bands[47,48].

**Methods:**

**Device fabrication**

Few-layer WSe$_2$, graphene and h-BN flakes were mechanically exfoliated from bulk crystals onto SiO$_2$ (285 nm)/Si substrates. The layer numbers of graphene and WSe$_2$ were initially identified from their optical contrasts and later confirmed by Raman spectroscopy, photoluminescence, and atomic force microscopy (AFM). To assemble the van der Waals heterostructures, we utilized the cut-and-stack technique. Bilayer WSe$_2$ flakes were cut into two parts using a sharp tungsten tip. The heterostructures were assembled in the sequence of top h-BN/bilayer graphene/twisted double bilayer WSe$_2$/bottom h-BN/bottom graphite using a standard dry transfer technique at temperatures between 90 °C and 120 °C with the assistance of a poly(bisphenol A carbonate) (PC)/polydimethylsiloxane (PDMS) stamp. During the assembly of bilayer WSe$_2$, we carefully picked up the first part of the pre-cut flake, then intentionally rotated the remaining part on the stage to achieve a target twist angle of 0.5°-2°, and subsequently picked up the second part of the flake. The final stacks were imaged using AFM, and bubble-free regions were selected as the channel area of the devices to prevent inhomogeneity and strain. One-dimensional electrical contacts to the graphene were achieved by dry etching with CHF$_3$/O$_2$ plasma and the deposition of Cr/Au (3/60 nm). The top gate was defined by electron-beam lithography (EBL), followed by the deposition of Cr/Au (3/50 nm). The final Hall bar was shaped through additional steps of EBL and etching process.

**PFM measurements**

The samples for PFM characterizations were prepared using a similar process to that for transport devices, with a slight modification. Few-layer graphite and twisted bilayer MoS$_2$ flakes were subsequently picked up using a PC/PDMS stamp. The PC film with the stack was exfoliated from the PDMS and flip onto fresh PDMS. The PC film was then dissolved in N-Methyl-2-pyrrolidone (NMP) solution. Finally, the stack was transferred onto another substrate. The twisted MoS$_2$ flakes prepared by this method were clean enough for PFM characterization.

The vertical PFM measurements were conducted with the Asylum Research Cypher S AFM at room temperature. The tip used was Muti75E-G, with a force constant of around 3 N m$^{-1}$. The measured contact resonance frequency was around 340 kHz. The applied a.c. bias voltage was 1 V.

**Transport Measurements**

We carried out the transport measurements in a cryostat (Oxford TeslatronPT) at a base temperature of 1.6 K, equipped with a superconducting magnet. Standard low-frequency lock-in techniques were used to acquire the data, with an a.c. excitation bias of 100 nA. The temperature-dependent measurements were performed using the VTI temperature controller.

We identified the twist angle of WSe$_2$ from the corresponding carrier density at which pronounced resistance peaks symmetrically appearing at both sides of graphene's CNP. We assigned them as the full filling carrier density $n_s(\nu = \pm 4) = 4/A$ by considering 2-fold spin and 2-fold valley degeneracy in graphene, where $A$ is the unit cell area of moiré superlattice. The moiré superlattice is created from parallelly twisting WSe$_2$ with an angle of $\theta$, giving $A = \sqrt{3}a^2/8\sin^2(\theta/2)$, where $a = 0.3297$ nm is the lattice constant of WSe$_2$.

The dual-gate configuration allows us to independently tune the carrier density $n$ and displacement field $D$ through $n = (C_b V_b + C_t V_t)/e$ and $D = (C_b V_b - C_t V_t)/2\varepsilon_0$, where $C_b$ ($C_t$) is the back (top) gate capacitance per unit area, $V_b$ ($V_t$) is the back (top) gate voltage, $e$ is an elementary charge and $\varepsilon_0$ is the vacuum permittivity. $C_b$ and $C_t$ were extracted through Hall density measurement at $\pm 1$ T with anti-symmetrized treatment to remove longitudinal components.

**Crystallographic alignment of graphene and h-BN in Device D2**

The existence of graphene/h-BN moiré superlattices can be further confirmed from optical images and Raman spectroscopy besides the transport results. Figure S2a shows the optical image of the final stack for Device D2. Obviously, the straight edge of graphene flake was aligned with the straight edge of the top h-BN. Raman spectrum gives more persuasive evidence. Figure S2b shows the 2D peaks of the Raman spectrum acquired from the stack containing bilayer graphene region. The peaks can be fitted by four sub-peaks, whose full width at half maximum (FWHM) yields a moiré wavelength $\lambda > 11$ nm according to previous statistical data[49]. We also acquired a Raman spectrum from an adjacent monolayer graphene, which shares the same crystallographic orientation as that of bilayer graphene. The 2D peak can be fitted by a single Lorentz function with FWHM of 33 cm$^{-1}$, exhibiting a typical single alignment feature[50].

**References:**

1   Yankowitz, M. *et al.* Emergence of superlattice Dirac points in graphene on hexagonal boron nitride. *Nat. Phys.* **8**, 382-386 (2012).
2   Dean, C. R. *et al.* Boron nitride substrates for high-quality graphene electronics. *Nat. Nanotechnol.* **5**, 722-726 (2010).
3   Wang, L. *et al.* One-dimensional electrical contact to a two-dimensional material. *Science* **342**, 614-617 (2013).
4   Zhou, H. *et al.* Half- and quarter-metals in rhombohedral trilayer graphene. *Nature* **598**, 429-433 (2021).
5   Zhou, H., Xie, T., Taniguchi, T., Watanabe, K. & Young, A. F. Superconductivity in rhombohedral trilayer graphene. *Nature* **598**, 434-438 (2021).
6   Shi, Y. *et al.* Electronic phase separation in multilayer rhombohedral graphite. *Nature* **584**, 210-214 (2020).
7   Chen, G. *et al.* Evidence of a gate-tunable Mott insulator in a trilayer graphene moiré superlattice. *Nat. Phys.* **15**, 237-241 (2019).

8   Chen, G. *et al.* Tunable correlated Chern insulator and ferromagnetism in a moiré superlattice. *Nature* **579**, 56-61 (2020).
9   Chen, G. *et al.* Signatures of tunable superconductivity in a trilayer graphene moiré superlattice. *Nature* **572**, 215-219 (2019).
10  Novoselov, K. S., Mishchenko, A., Carvalho, A. & Castro Neto, A. H. 2D materials and van der Waals heterostructures. *Science* **353**, aac9439 (2016).
11  Ponomarenko, L. A. *et al.* Cloning of Dirac fermions in graphene superlattices. *Nature* **497**, 594-597 (2013).
12  Dean, C. R. *et al.* Hofstadter's butterfly and the fractal quantum Hall effect in moiré superlattices. *Nature* **497**, 598-602 (2013).
13  Hunt, B. *et al.* Massive Dirac fermions and Hofstadter butterfly in a van der Waals heterostructure. *Science* **340**, 1427-1430 (2013).
14  Cao, Y. *et al.* Correlated insulator behaviour at half-filling in magic-angle graphene superlattices. *Nature* **556**, 80-84 (2018).
15  Cao, Y. *et al.* Unconventional superconductivity in magic-angle graphene superlattices. *Nature* **556**, 43-50 (2018).
16  Sharpe, A. L. *et al.* Emergent ferromagnetism near three-quarters filling in twisted bilayer graphene. *Science* **365**, 605-608 (2019).
17  Zheng, Z. *et al.* Unconventional ferroelectricity in moire heterostructures. *Nature* **588**, 71-76 (2020).
18  Lu, Z. *et al.* Fractional quantum anomalous Hall effect in multilayer graphene. *Nature* **626**, 759-764 (2024).
19  Serlin, M. *et al.* Intrinsic quantized anomalous Hall effect in a moiré heterostructure. *Science* **367**, 900-903 (2020).
20  Yasuda, K., Wang, X., Watanabe, K., Taniguchi, T. & Jarillo-Herrero, P. Stacking-engineered ferroelectricity in bilayer boron nitride. *Science* **372**, 1458-1462 (2021).
21  M. Vizner Stern, Y. W., W. Cao,I.Nevo, K. Watanabe, T. Taniguchi, E. Sela, M. Urbakh,O. Hod, M. Ben Shalom. Interfacial ferroelectricity by van der Waals sliding. *Science* **372**, 1462-1466 (2021).
22  Woods, C. R. *et al.* Charge-polarized interfacial superlattices in marginally twisted hexagonal boron nitride. *Nat. Commun.* **12**, 347 (2021).
23  Wang, X. *et al.* Interfacial ferroelectricity in rhombohedral-stacked bilayer transition metal dichalcogenides. *Nat. Nanotechnol.* (2022).
24  Weston, A. *et al.* Interfacial ferroelectricity in marginally twisted 2D semiconductors. *Nat. Nanotechnol.* **17**, 390-395 (2022).
25  Deb, S. *et al.* Cumulative polarization in conductive interfacial ferroelectrics. *Nature* **612**, 465-469 (2022).
26  Ko, K. *et al.* Operando electron microscopy investigation of polar domain dynamics in twisted van der Waals homobilayers. *Nat. Mater.* **22**, 992-998 (2023).
27  Fei, Z. *et al.* Ferroelectric switching of a two-dimensional metal. *Nature* **560**, 336-339 (2018).
28  Wan, Y. *et al.* Room-Temperature Ferroelectricity in 1T′-ReS$_2$ Multilayers. *Phys. Rev. Lett.* **128** (2022).
29  Li, L. & Wu, M. Binary Compound Bilayer and Multilayer with Vertical

Polarizations: Two-Dimensional Ferroelectrics, Multiferroics, and Nanogenerators. *ACS Nano* **11**, 6382-6388 (2017).

30  Bennett, D. & Remez, B. On electrically tunable stacking domains and ferroelectricity in moiré superlattices. *npj 2D Mater. Appl.* **6** (2022).

31  Bennett, D. Theory of polar domains in moiré heterostructures. *Phys. Rev. B* **105** (2022).

32  Zhao, P., Xiao, C. & Yao, W. Universal superlattice potential for 2D materials from twisted interface inside h-BN substrate. *npj 2D Mater. Appl.* **5** (2021).

33  Gu, J. *et al.* Remote imprinting of moiré lattices. *Nat. Mater.* **23**, 219-223 (2024).

34  Zhang, Z. *et al.* Engineering correlated insulators in bilayer graphene with a remote Coulomb superlattice. *Nat. Mater.* **23**, 189-195 (2024).

35  Zhang, S. *et al.* Visualizing moiré ferroelectricity via plasmons and nano-photocurrent in graphene/twisted-$WSe_2$ structures. *Nat. Commun.* **14**, 6200 (2023).

36  Kim, D. S. *et al.* Electrostatic moiré potential from twisted hexagonal boron nitride layers. *Nat. Mater.* **23**, 65-70 (2024).

37  Kim, K. *et al.* Band Alignment in $WSe_2$–Graphene Heterostructures. *ACS Nano* **9**, 4527-4532 (2015).

38  Island, J. O. *et al.* Spin-orbit-driven band inversion in bilayer graphene by the van der Waals proximity effect. *Nature* **571**, 85-89 (2019).

39  Weston, A. *et al.* Atomic reconstruction in twisted bilayers of transition metal dichalcogenides. *Nat. Nanotechnol.* **15**, 592-597 (2020).

40  McGilly, L. J. *et al.* Visualization of moire superlattices. *Nat. Nanotechnol.* **15**, 580-584 (2020).

41  Woods, C. R. *et al.* Commensurate–incommensurate transition in graphene on hexagonal boron nitride. *Nat. Phys.* **10**, 451-456 (2014).

42  Kim, M. *et al.* Control of electron-electron interaction in graphene by proximity screening. *Nat. Commun.* **11**, 2339 (2020).

43  Wang, Z. *et al.* Composite super-moiré lattices in double-aligned graphene heterostructures. *Sci. Adv.* **5**, eaay8897 (2019).

44  Zhou, B., Yang, H. & Zhang, Y.-H. Fractional quantum anomalous Hall effects in rhombohedral multilayer graphene in the moiréless limit and in Coulomb imprinted superlattice. *arXiv*, arXiv:2311.04217 (2023).

45  Dong, Z., Patri, A. S. & Senthil, T. Theory of fractional quantum anomalous Hall phases in pentalayer rhombohedral graphene moiré structures. *arXiv*, arXiv:2311.03445 (2023).

46  Dong, J. *et al.* Anomalous Hall Crystals in Rhombohedral Multilayer Graphene I: Interaction-Driven Chern Bands and Fractional Quantum Hall States at Zero Magnetic Field. *arXiv*, arXiv:2311.05568 (2023).

47  Ghorashi, S. A. A. *et al.* Topological and Stacked Flat Bands in Bilayer Graphene with a Superlattice Potential. *Phys. Rev. Lett.* **130**, 196201 (2023).

48  Zeng, Y. *et al.* Gate-tunable topological phases in superlattice modulated bilayer graphene. *Phys. Rev. B* **109**, 195406 (2024).

49  Cheng, B. *et al.* Raman spectroscopy measurement of bilayer graphene's twist angle to boron nitride. *Appl. Phys. Lett.* **107**, 033101 (2015).

50  Finney, N. R. *et al.* Tunable crystal symmetry in graphene-boron nitride heterostructures with coexisting moire superlattices. *Nat. Nanotechnol.* **14**, 1029-1034 (2019).

**Acknowledgements**

This work was funded by National Natural Science Foundation of China (Grant No. 12274354), the Zhejiang Provincial Natural Science Foundation of China (Grant No. LR24A040003; XHD23A2001), and Westlake Education Foundation at Westlake University. We thank Chao Zhang from the Instrumentation and Service Center for Physical Sciences (ISCPS) at Westlake University for technical support in data acquisition. We also thank Westlake Center for Micro/Nano Fabrication and the Instrumentation and Service Centers for Molecular Science for facility support. K.W. and T.T. acknowledge support from the JSPS KAKENHI (Grant Numbers 21H05233 and 23H02052) and World Premier International Research Center Initiative (WPI), MEXT, Japan.

**Author Contributions**

S.X. conceived the idea and supervised the project. J.D. fabricated the devices with the assistance of N.L., X.F., K.W., and L.W.. N.L. grew the $WSe_2$ crystals. J.D. performed the transport measurements with the assistance of W.Z.. H.X. performed AFM measurements. J.D. and H.X. performed Raman measurements. K.W. and T.T. grew h-BN crystals. J.D. and S.X. analyzed data and wrote the paper. All the authors contributed to the discussions.

# Figures

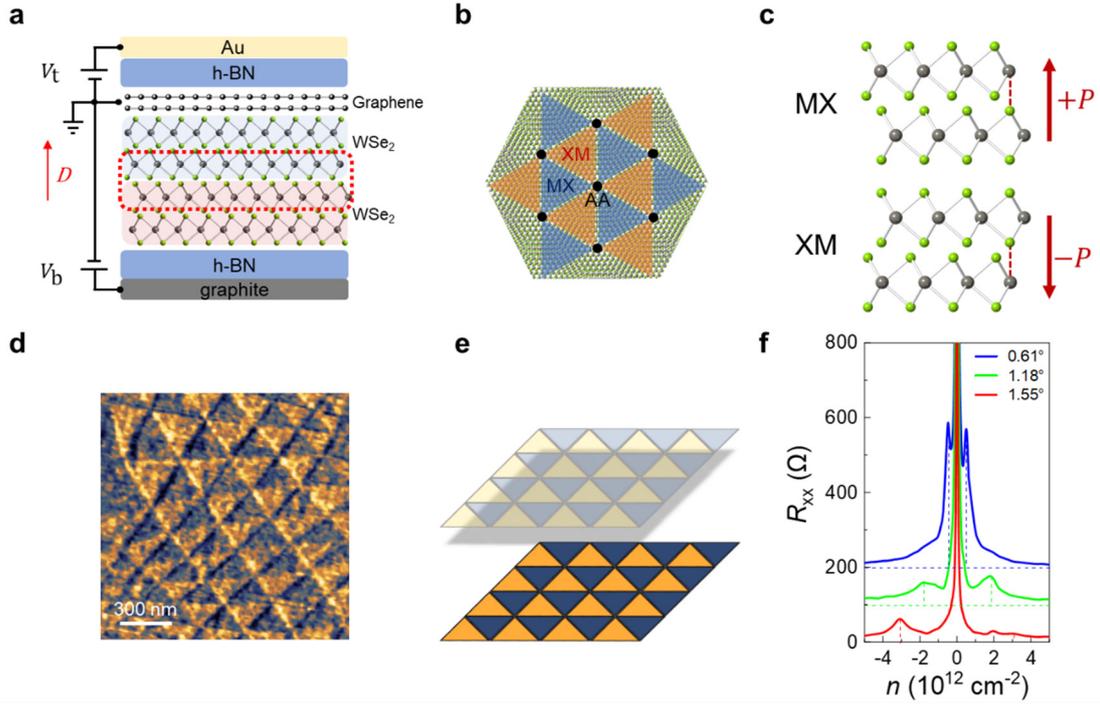

**Fig. 1: Tunable superlattices in bilayer graphene imprinted by remote moiré ferroelectricity. a**, Schematic of the device structure. The blue and red shadow areas mark the top and bottom bilayer $WSe_2$, respectively. The red dashed box denotes the parallel-stacked interface. **b**, Schematic of domain configurations in parallel-stacked TMDCs. Two kinds of rhombohedral-stacked (MX and XM) domains dominate in the system arising from lattice reconstructions. **c**, The vertical alignments of M and X atoms for MX and XM domains. The rhombohedral stackings endow MX and XM domains with out-of-plane polarizations, but opposite directions. **d**, Vertical PFM image of a parallel-stacked $MoS_2$. **e**, Illustration of the process of a ferroelectric moiré superlattice imprinting its potential onto a remote bilayer graphene. This strategy separates the moiré construction layer from the electronic transport layer. **f**, Longitudinal resistance $R_{xx}$ as a function of carrier density $n$ measured at $T = 1.6$ K for Device D1 (60°-0.61°), Device D3 (180°-1.18°), Device D4-1 (60°+1.55°). Two satellite resistance peaks symmetrically appear near CNP, suggesting that the constructed moirés bands in bilayer graphene are tunable by the twist angle of $WSe_2$.

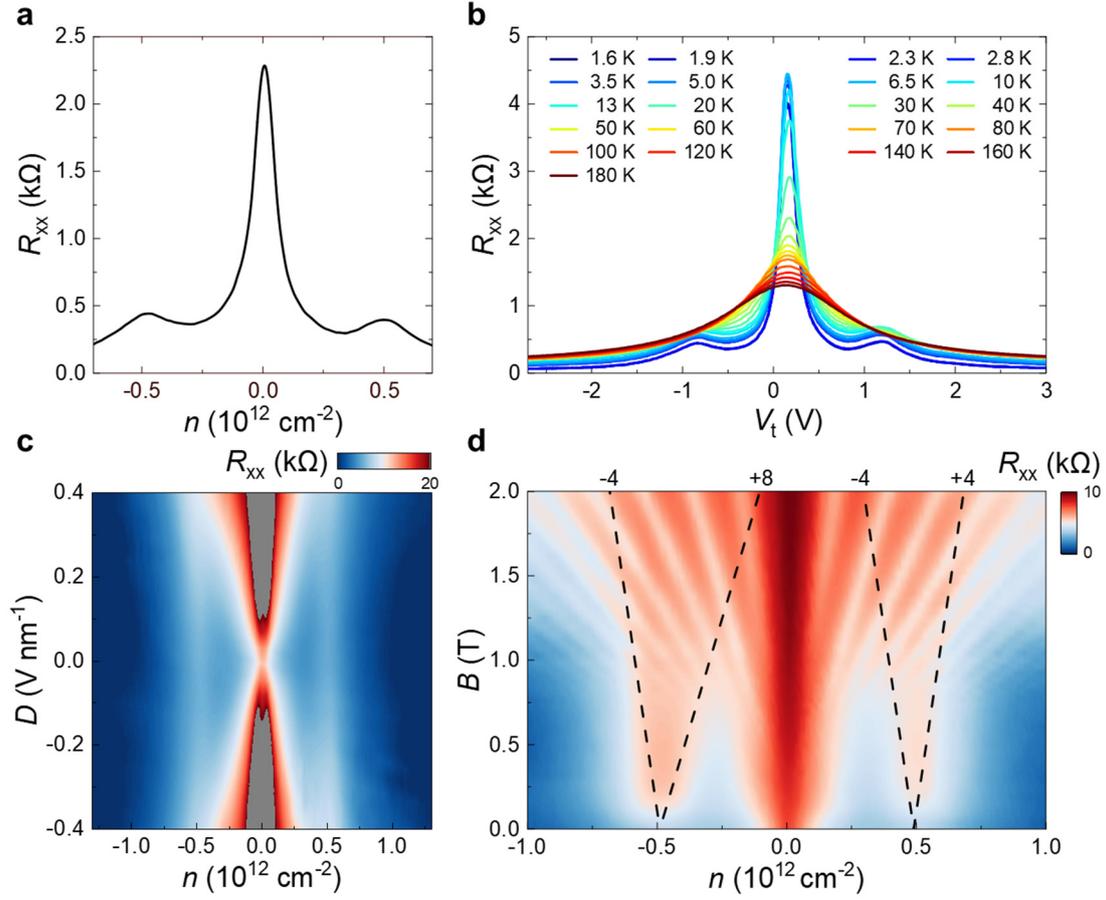

**Fig. 2: Band structure engineering of bilayer graphene by twisted double bilayer WSe$_2$ with a twist angle of 60°+0.61° (Device D1). a**, $R_{xx}$ as a function of $n$ measured at $D = 0$ V nm$^{-1}$. **b**, Temperature dependent $R_{xx}$ as a function of $V_t$ at a fixed $V_b = 0$ V. **c**, The color plot of the $n-D$ map of $R_{xx}$. **d**, The color plot of $R_{xx}$ as a function of $n$ and $B$ at a fixed $D = 0$ V nm$^{-1}$. The black dashed lines mark the Landau levels with corresponding filling factors. The data in (**a**), (**c**), and (**d**) were measured at $T = 1.6$ K.

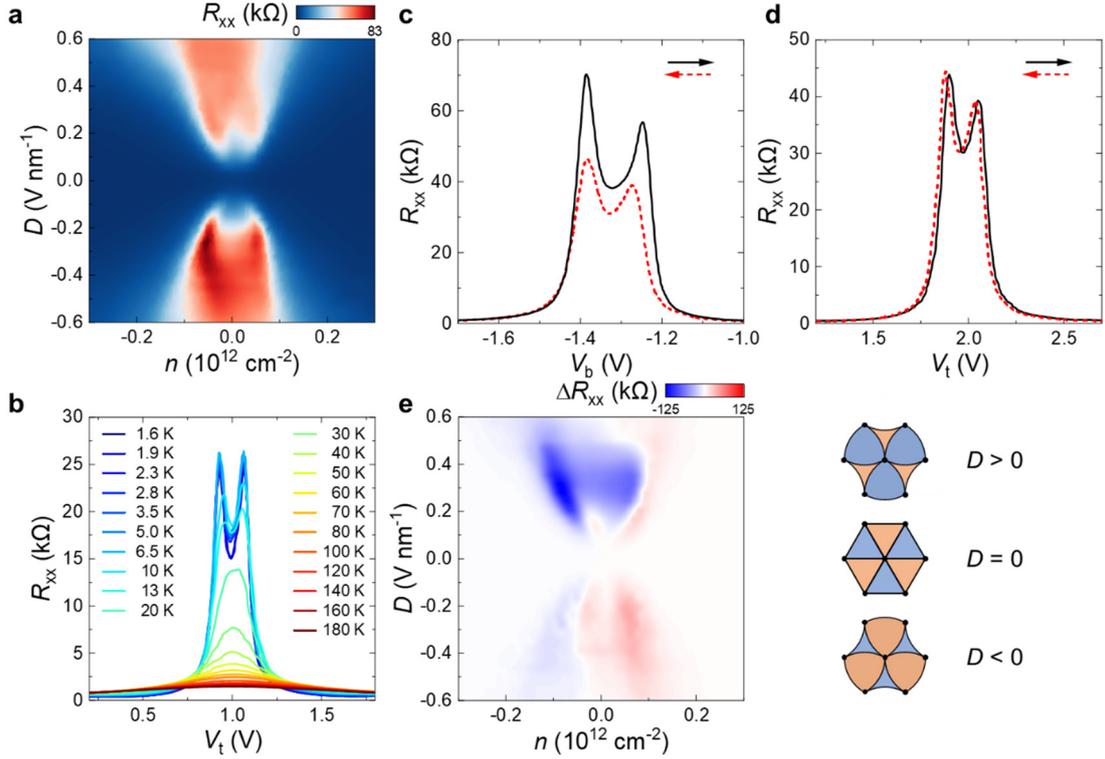

**Fig. 3: Ferroelectric hysteresis at finite displacement fields. a**, The enlarged plot of the $n - D$ map of $R_{xx}$ near CNP. **b**, Temperature dependent $R_{xx}$ as a function of $V_t$ at a fixed $V_b = -0.6$ V. **c**, $R_{xx}$ as a function of $V_b$ by sweeping $V_b$ forward (black solid line) and backward (red dashed line) at a fixed $V_t = 2$ V. **d**, $R_{xx}$ as a function of $V_t$ by sweeping $V_t$ forward (black solid line) and backward (red dashed line) at a fixed $V_b = -1.3$ V. **e**, The color plot of the difference in $R_{xx}$ between forward and backward sweeps of $V_b$ at each fixed $V_t$. During this measurement, $V_b$ is the fast-scan axis and $V_t$ is the slow-scan axis. The $R_{xx}(V_b, V_t)$ is converted to $R_{xx}(n, D)$ using the way described in the Methods. All the data were measured in Device D1.

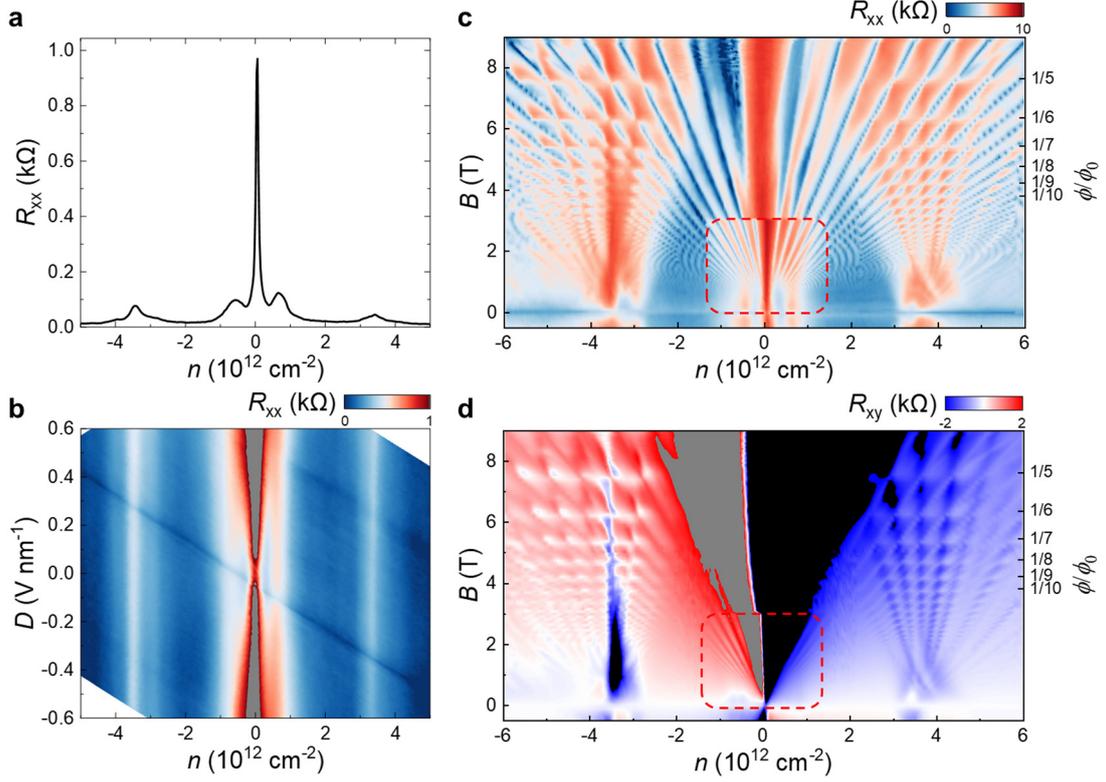

**Fig. 4: Coexistence of ferroelectric moiré and conventional moiré superlattices in Device D2. a**, $R_{xx}$ as a function of $n$ measured at $D = 0$ V nm$^{-1}$. Two sets of satellite peaks symmetrically appear near CNP. **b**, The color plot of the $n - D$ map of $R_{xx}$, showing two groups of vertical lines besides the CNP. **c, d**, The color plot of $R_{xx}$ (**c**) and $R_{xy}$ (**d**) as a function of $n$ and $B$ at a fixed $D = 0$ V nm$^{-1}$. The red dashed boxes show the fine maps of Landau fan diagrams near ferroelectric moiré induced resistance peaks. The right $y$ axis shows the normalized magnetic flux $\phi/\phi_0$, where the commensurate states $(1/q)$ are marked. All the data were measured at $T = 1.6$ K.

**Tab. 1: List of measured devices**

| Device | Twisted layer of $WSe_2$ | $n_s$ ($10^{12}$ cm$^{-2}$) | λ (nm) | Angle |
|---|---|---|---|---|
| D1 | 2 + 2 | 0.49 | 30.8 | 60° - 0.61° |
| D2 | 1 + 1 | 0.65 | 26.6 | 0.71° ($WSe_2$) |
|  |  | 3.46 ($n_s$') | 11.5 | 0.81° (h-BN/graphene) |
| D3 | 2 + 2 | 1.80 | 16.0 | 180° - 1.18° |
| D4-1 | 2 + 2 | 3.11 | 12.2 | 60° + 1.55° |
| D4-2 | 2 + 2 | 2.64 | 13.2 | 60° + 1.43° |

*All the angles were targeted to achieve parallel-stacked interfaces, namely around 60° (or 180°) for twisted double bilayer $WSe_2$ and around 0° for twisted bilayer $WSe_2$.

**The sign of the twist angle $\theta$ is assigned according to the target angle during the transfer.